# The fast spin-rotation of a young extra-solar planet

Ignas A. G. Snellen[1], Bernhard R. Brandl[1], Remco J. de Kok[1,2], Matteo Brogi[1], Jayne Birkby[1], and Henriette Schwarz[1]

[1]*Leiden Observatory, Leiden University, Postbus 9513, 2300 RA Leiden, The Netherlands*

[2]*SRON, Sorbonnelaan 2, 3584 CA Utrecht, The Netherlands*

**The spin-rotation of a planet arises from the accretion of angular momentum during its formation[1,2,3], but the details of this process are still unclear. In the solar system, the equatorial rotation velocities and spin angular momentum of the planets show a clear trend with mass[4], except for Mercury and Venus which have significantly spun down since their formation due to tidal interactions[5,6]. Here we report on near-infrared spectroscopic observations at $\lambda/\Delta\lambda$=100,000 of the young extra-solar gas giant β Pictoris b[7,8]. The absorption signal from carbon monoxide in the planet's thermal spectrum is found to be blueshifted with respect to the velocity of the parent star by (-15±1.7) km sec$^{-1}$, consistent with a circular orbit[9]. The combined line profile exhibits a rotational broadening of 25±3 km sec$^{-1}$, meaning that β Pictoris b spins significantly faster than any planet in the solar system, in line with the extrapolation of the known trend in spin velocity with planet mass.**

Near-infrared high-dispersion spectroscopy has been used to characterize the atmospheres of hot Jupiters in close-in orbits[10-11]. Such observations utilize changes in the radial component of the orbital velocity of the planet (resulting in changes in Doppler shift) to filter out the quasi-stationary telluric and stellar contributions in the spectra. Here we make use of the spatial separation and the difference in spectral signature between the planet and star, which allow the starlight to be filtered out. A similar technique[12] has been applied very successfully[13] at a medium spectral dispersion to characterize the exoplanet HR8799c. We observed the β Pictoris system[7,8] (K=3.5) using the Cryogenic high-Resolution InfraRed Echelle Spectrograph CRIRES[14] located at the Nasmyth focus of UT1 of the Very Large Telescope (VLT) of the European Southern Observatory (ESO) at Cerro Paranal in Chile on the night of 17 December 2013, with the slit oriented in such way that it encompassed the planet and star.

An important step in the data analysis is the optimal removal of the stellar contribution along the slit, which for this A-star consists mostly of a telluric absorption spectrum. The resulting spectra were cross-correlated with theoretical spectral templates constructed in a similar way as in our previous work on hot Jupiters[10-11], varying the planet's atmospheric temperature pressure (T/p) profile, the carbon monoxide abundance and that of water vapor and methane, which can also show features in the observed wavelength range. Note that there is a strong degeneracy between the atmospheric T/p profile and the abundance of the molecular species, meaning that different combinations of these parameters result in nearly identical template spectra.

At the expected planet position a broad and blue-shifted signal is apparent (see Figure 1), which is strongest when the cross-correlation is performed with a spectral template from an atmospheric model with deep carbon monoxide lines and a small contribution from water. We estimate the signal to have a SNR of 6.4 by cross-correlating the residual spectrum with a broadened model template, and compare the peak of the cross-correlation profile with the standard deviation. If we use the cross-correlation profile as seen in Figure 1 to estimate the SNR, we need to take into account the width of the signal and the dependence of adjacent pixels in the profile. This results in an SNR of 7.8, but this latter method we found to be less accurate, since it does not properly include contributions from correlated noise structures on scales of the broad signal. Cross-correlation with the optimal spectrum of water vapor alone provides a marginal signal at SNR~2, which means we cannot claim a firm detection of water in the planet atmosphere. No signal is retrieved for methane models (see Extended Data Fig. 1).

We fit the planet profile using a grid of artificial cross-correlation functions, produced by cross-correlating the optimal template spectrum with a broadened and velocity-shifted copy of itself, for a range of velocities and rotational broadening functions. The best fit (left panel of Figure 2) was obtained for a blue-shifted radial velocity of -15.4±1.7 km sec$^{-1}$ (1$\sigma$, including the uncertainties in the system velocity, +20±0.7 km sec$^{-1}$, of the star[15]) and a rotational broadening of 25±3 km sec$^{-1}$ (1$\sigma$), with the uncertainties estimated from chi-squared analysis. Cross-correlation with other less optimal model templates give only small variations in these values at the < 0.3$\sigma$ level (see Methods section). Monitoring the position of the planet over the last decade[9] has resulted in an estimate of the orbital radius of 8 - 9 AU and an orbital period of 17-20 years. The planet ephemeris[9], assuming a circular orbit, provides a planet radial velocity relative to the star of 13.2 - 14.1 km sec$^{-1}$ (1$\sigma$) either towards or away from us. This means there is no evidence for an eccentric orbit from our one planet radial velocity measurement. Since the orbital inclination has been constrained[9] to >86.8º, the planet has a ~20% probability to transit its parent star. Marginal evidence was presented for a possible transit event[16,17] in 1981. That the planet is now moving towards us means that it is approaching inferior conjunction in the period 2017.5-2018.5, and previously during 1998-2001 and 1978-1984. This is consistent with a transit event in 1981. Note that if this event is spurious, there was an a priori probability of 50% for the direction (e.g. clockwise or anti-clockwise) of the planet motion to be consistent with the 1981 event.

The equatorial spin-rotation velocity of $V_{spin}$ = 25 km sec$^{-1}$ of β Pictoris b is higher than for any planet in the Solar-System. The Solar System planets show a clear correlation between their equatorial rotation velocity (or spin angular momentum) and mass[4] (see the right panel of Figure 2), the former ranging from 0.24 km sec$^{-1}$ for Mars to 13.3 km sec$^{-1}$ for Jupiter. Both Mercury and Venus have dropped significantly below this relation since they have spun down from the tidal interactions with the Sun[5,6]. Also the Earth has slowed down due to the gravitational interaction with the Moon[18]. The origin of the $V_{spin}$-mass relation is not known, but must be linked to the mass accretion processes during planet formation. The mass of β Pictoris b is highly uncertain, with estimates depending strongly on the time since its formation. A commonly used age estimate for the β Pictoris system is ($12^{+8}_{-4}$) Myr, resulting in mass estimates[19,20] of ($7^{+4}_{-3}$) $M_{Jup}$ to ($10^{+3}_{-2}$) $M_{Jup}$. However, recently, lithium depletion in the β Pictoris moving group[21] suggests an age of 21±4 Myr, which would increase these planet mass estimates by ~30%. We therefore adopt a planet mass of 11±5 $M_{Jup}$. As can be seen in panel b of Figure 2, β Pictoris b appears broadly consistent with an extrapolation of the solar system $V_{spin}$-mass relation to higher planet masses. However, the spread in measured projected spin velocities of L4-L7 brown dwarfs[22,23] (open circles) suggest a more complicated relationship for the general population of substellar objects. Furthermore, despite rotating faster than any planet in the solar system, β Pictoris b does appear to have a lower $V_{spin}$ than expected from a direct extrapolation of the solar system trend, which would point to a spin velocity of 50±12 km sec$^{-1}$. This may well be because the planet is still young and warm (1575-1650 K) and therefore bloated. Although the radius of the planet cannot be measured directly as in the case of transiting planets, spectral modeling[19] of broadband photometry obtained by direct imaging provides an estimate of 1.65±0.06 $R_{Jup}$. It implies a rotation period, the length of day on β Pictoris b, of ~8.1±1.0 hours. Over the next hundreds of millions of years the planet is expected to cool down and shrink to the size of Jupiter[24]. If its angular moment is preserved during this process, the planet should spin up to $V_{spin}$ ~ 40 km sec$^{-1}$, depending on possible changes in its internal structure, decreasing its length of day to ~3 hours. The reader should note that in the above we assume that the obliquity, the axial tilt of the planet, is small – which may not be the case, as we know from Uranus. However, even if we assume that the distribution of obliquities of extra-solar gas giants is random, the average component in the direction of Earth is 30°, resulting in an underestimate of $V_{spin}$ of only 1 / cos(30°) = 15%.

**Methods Summary**

We targeted the 2-0 R-branch of carbon monoxide absorption using the two central detectors of CRIRES, resulting in a wavelength range of 2.303 to 2.331 μm. The CRIRES slit (width=0.2″) was oriented at 30° North through East to encompass the planet, which, according to recent high-contrast imaging observations, was located 0.4″ South-West from the host star[19,20]. We utilized the Multi-Application Curvature Adaptive Optics system (MACAO)[25] in 1-1.3″ seeing conditions. This resulted in a reduction of the starlight at the planet position of a factor 8 to 30 (depending on the seeing). Therefore, a certain planet/star contrast can be reached 8-30 times faster compared to 'classical' high-dispersion spectroscopy of hot Jupiters, because the noise in the final planet spectrum originates from the star. After basic calibrations, we optimally extracted a 1D spectrum for each position along the slit, and subsequently removed the stellar contribution for each position. The resulting spectra were cross-correlated with planet model spectra (Extended Data Fig.1), which were produced in a similar way as for our previous work on hot Jupiters[10-12].

The SNR that can be achieved on a planet spectrum for this type of observation is a strong function of telescope diameter D. This opens the way of obtaining two-dimensional maps of the planet using Doppler imaging, a technique used to map spot distributions on fast-rotating active stars[26-29]. Very recently, a first Doppler image map was produced for the nearby brown dwarf Luhman 16B (K=9.73) using CRIRES on the VLT, showing large-scale bright and dark features, indicative of patchy clouds[23]. The planet β Pictoris b is only a factor 13 fainter than Luhman 16B. Our simulations (Extended Data Fig.2) show that a similar study can be conducted on β Pictoris b using future instrumentation.

**Acknowledgements** We thank ESO director Tim de Zeeuw for granting Director's Discretionary Time on the Very Large Telescope to perform these observations (292.C-5017(A)). I.S. acknowledges support from an NWO VICI grant. R.d.K acknowledges the NWO PEPSci program.

**Author Contributions** I.S. conceived the project with help from B.B, R.d.K, M.B, J.B. The analysis was led by I.S. and he wrote the first version of the manuscript. I.S. and B.B. conceived the connection with the ELT. R.d.K constructed the planet atmosphere models. B.B, R.d.K, M.B., J.B., and H.S. discussed the analyses, results, and commented on the manuscript.

**Author Information** The authors declare no competing financial interests. Correspondence and request for materials should be addressed to I.S. (e-mail: snellen@strw.leidenuniv.nl).

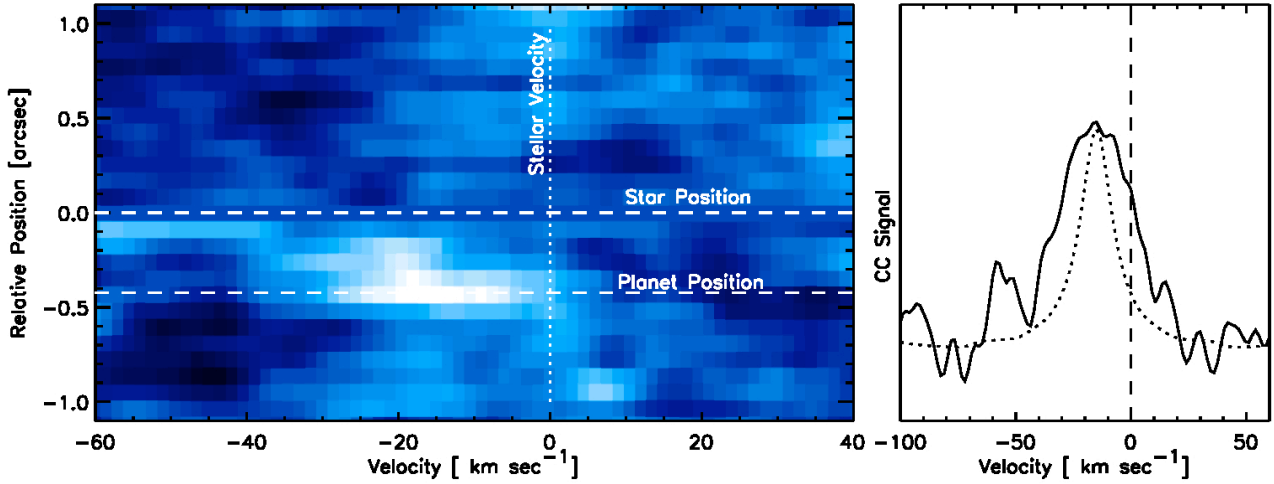

**Figure 1. The broadened cross-correlation signal of β Pictoris b.** Panel a) shows the CO+$H_2O$ cross-correlation (CC) signal as function of the position along the slit (rotated by 30° North to East), after the stellar contribution was removed. The x-axis shows the radial velocity with respect the system velocity (+20±0.7 km sec$^{-1}$) of the star[15]. The y-axis denotes the relative position with respect to the star β Pictoris with the planet located 0.4″ below, both indicated by horizontal dashed lines. A broad signal, at an SNR of 6.4 is visible, blue-shifted by -15.4±1.7 km sec$^{-1}$ (1σ) with respect to the system velocity. Panel b) shows the CC signal at the planet position. The dotted curve shows the arbitrarily scaled auto-correlation function of the λ/Δλ=100,000 model template, indicating the CC-signal expected from a non-rotating planet.

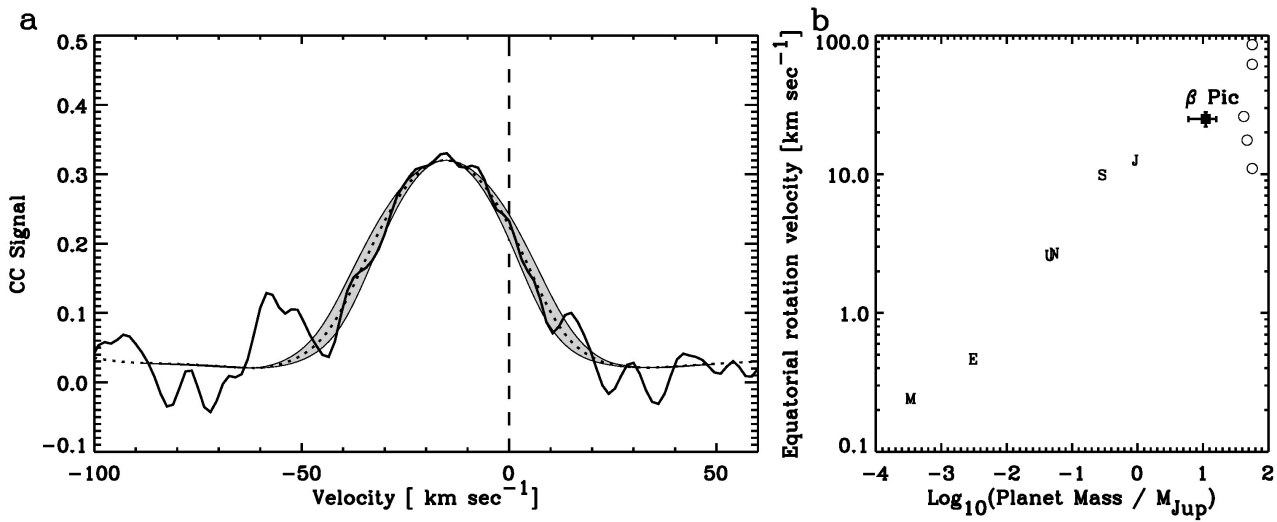

**Figure 2. The Spin-rotation of β Pictoris b.** Panel a) shows the $CO+H_2O$ cross correlation signal of the planet β Pictoris b (solid line). The dashed line indicates the best fit to the cross-correlation profile of the model template with a copy of itself, rotationally broadened by 25 km sec$^{-1}$. The grey area indicates the 1σ uncertainty in the level of broadening of ±3 km sec$^{-1}$. Panel b) shows the equatorial rotation velocity of the solar system planets as function of planet mass, each planet indicated by its first letter. Note that Mercury and Venus are not visible on this plot. Tidal interactions with the Sun made these planets to significantly spin down to $V_{spin}$ <<0.1 km sec$^{-1}$. β Pictoris b is indicated with a planet mass of 11$^{\pm}$5 $M_{Jup}$, consistent with the solar system mass-$V_{spin}$ trend extended to higher masses. The open circles indicate the projected spin velocities of L4-L7 type brown dwarfs from refs 22 and 23.

## Methods

**Observations and Data Analysis**

We targeted the 2-0 R-branch of carbon monoxide absorption using the two central detectors of CRIRES, resulting in a wavelength range of 2.303 to 2.331 μm. The slit (width=0.2″) was oriented at 30° North through East to encompass the planet, which, according to recent high-contrast imaging observations, was located 0.4″ South-West from the host star[19,20]. We took 44 exposures of 4x10 seconds utilizing an ABBA dither pattern to enable accurate background subtraction. In addition, each dither position was given a small but random offset to minimize potential flat-fielding issues. We utilized the Multi-Application Curvature Adaptive Optics system (MACAO)[25] in 1-1.3″ seeing conditions. This resulted in a reduction of the starlight at the planet position of a factor 8 to 30 (depending on the seeing). Therefore, a certain planet/star contrast can be reached 8-30 times faster compared to 'classical' high-dispersion spectroscopy of hot Jupiters, because the noise in the final planet spectrum originates from the star.

We performed the basic calibrations on the individual frames using the standard CRIRES reduction pipeline[30], including dark subtraction, flat fielding, and non-linearity corrections, after which the frames in the AB or BA pairs of two points along the slit were subtracted from each other. Subsequently, the frames of each pair of two-dimensional spectra were combined. The subsequent analysis was performed on the resulting 22 intermediate data products from the pipeline using IDL routines. We determined the spatial profiles of the star as function of wavelength, which we subsequently used to optimally extract a 1D spectrum for each position along the slit, in steps of one spatial pixel of 0.086″, with the planet expected to be located about five pixels from the center of the star[9]. The combined spectrum of 30 pixel rows along the slit was taken as a reference of the stellar spectrum and used to remove the stellar contribution from each spatial pixel position along the slit – a crucial step in the data analysis. At the central position of the star, its spatial profile is narrow compared to the slit-width, resulting in a higher spectral resolution compared to elsewhere along the slit, in addition to a small wavelength offset because the star was not exactly centered in the slit. To remove the star light properly at each position along the slit, the reference spectrum not only had to be scaled, but also shifted and convolved with a broadening function. This broadening function was determined for each individual slit position and frame, using the Singular Value Decomposition technique[31]. The reader should note that the average reference spectrum contains planet light at a level of 1/8 to 1/30 of that at the planet position, marginally weakening the expected signal by that amount. The residual spectra as function of position along the slit for the 22 frames were subsequently combined, weighing the frames according to their varying (seeing induced) stellar profile, and thus maximizing the expected SNR of the final planet spectrum.

Planet model spectra were produced in a similar way as for our previous work on hot Jupiters[10,11,32,33]. A uniform temperature-Pressure (T/p) profile was assumed to describe the vertical structure of the planet atmosphere, parameterized by two points, $(T_1,p_1)$ and $(T_2,p_2)$. Below and above these levels in the atmosphere the temperature was assumed to be constant at $T_1$ and $T_2$, respectively. At pressures $p_1 > p > p_2$ a constant lapse rate was assumed as $\delta T/\delta \log_{10}p = (T_2-T_1)/(\log_{10}p_2-\log_{10}p_1)$. The model spectra were subsequently produced using line-by-line calculations including $H_2$-$H_2$ collision-induced absorption[34,35] and absorption from three trace gases, i.e. CO, $H_2O$, and $CH_4$, which are the dominant sources of opacities at these wavelengths[36,37]. Line data for the trace gases were taken[38,39] from HITEMP 2010 for $H_2O$ and CO, and HITRAN 2008 for $CH_4$. A Voigt line profile was used for the calculations.

Since large degeneracies exist between different models, we did not explore a full grid of models varying $(T_1,p_1)$, $(T_2,p_2)$ and the molecular abundances. This can be seen in Extended Data Figure 1, which shows the results of five different models, with the parameters indicated in the title. The two top models have a CO volume mixing ratio (VMR) that differ by three orders of magnitude ($VMR_{CO}=10^{-1.5}$ versus $10^{-4.5}$), but both show a strong signal. If we mix in other molecules, the significance of the detected signal becomes lower. This can be seen by cross-correlating with the third model, which is for a low CO and high $H_2O$ VMR, showing that the signal is still visible but at a significantly lower SNR than above. This is in line with the fourth model spectrum consisting of $H_2O$ only, which gives only a marginal detection at SNR~2. We therefore cannot claim a firm detection of water vapour in this planet atmosphere. The last model is for $CH_4$ only – showing no signal.

**Doppler Imaging simulations**

High-dispersion spectral observations that aim to characterize exoplanet atmospheres have so far been restricted to hot Jupiters in very close-in orbits. Such observations utilize changes in the radial component of the orbital velocity of the planet (resulting in changes in Doppler shift) to filter out the quasi-static telluric and stellar contributions in the spectra. Here we use high-dispersion spectroscopy to characterize the young gas giant β Pictoris b, which was previously discovered using high-contrast imaging techniques. These observations make use of the angular separation and the spectral differences between the planet and the star. While the planet to star contrast is about 3800 at 2.3 μm, the starlight is reduced by a factor 8 to 30 (varying with the seeing) at the planet position 0.4″ away, making the planet-light buried at a level 1/500th to 1/100th in the wing of the stellar spatial profile. This stellar contribution, which is dominated by telluric absorption, can be filtered out very effectively since this spectrum is well determined along the slit, albeit with small variations in the spectral dispersion which are taken into account, and it is therefore known at a very high signal to noise. A similar technique[12] has been applied very successfully[13] at a medium spectral dispersion to characterize the exoplanet HR8799c.

The relative ease at which the presented detection was obtained is noteworthy, using only 1 hour of observing time (including overheads) on an 8m-class telescope in mediocre seeing. If the CO signal was not smeared out over -25 to +25 km sec$^{-1}$ due to the planet rotation, but concentrated in an unresolved signal, it would have been detected at a SNR~15. We also investigated the scientific potential of the future European Extremely Large Telescopes (E-ELT) for these kinds of observations. METIS, the Mid-infrared E-ELT Imager and Spectrograph[40], will have a $\lambda/\Delta\lambda$=100,000 integral field spectrograph in the 3-5 μm wavelength range. This does not cover the 2.3 μm CO, but $H_2O$ at 3.2 or 3.5 μm can also be targeted highly effectively. The absorption features are probably somewhat weaker than those of CO at 2.3 μm, but both the planet/star contrast and the flux from the planet are more than a factor two higher. Alternatively, the HIRES spectrograph[41] is expected to cover the optical to near-infrared wavelength regime, and can be used for this type of observation, if it is designed with a slit or with multiple fibers that can cover both the star and the planet. The other Extremely Large Telescope projects, the Giant Magellan Telescope[42] and the Thirty Meter Telescope[43], also have optical and infrared high-resolution spectrographs in their instrument roadmaps.

In this work we focus on the E-ELT. For simplicity, we will assume a spectrograph concept similar to CRIRES, using an adaptive optics system that produces similar Strehl ratios under the same atmospheric conditions. Since the angular resolution of the telescope scales with $\lambda/D$, with D the telescope diameter (D=39m for the E-ELT), the angular distance of the planet from the star in units of diffraction elements will be a factor 4.8 larger. In our CRIRES frames the starlight is suppressed by a factor ~500 at that distance. In addition, the apparent size of the planet is decreased by a factor $D^2$, gaining another factor 23. This means that for similar observations with the E-ELT the starlight at the planet position will be reduced by a factor ~$10^4$, meaning it only marginally contributes to the noise in the planet spectrum. It implies that with the E-ELT we will be able to observe the K=12.5 planet as if there were no nearby host star. We use the CRIRES exposure time calculator, multiplying the number of observations by a factor 23 to account for the increase in collecting area, to estimate that an exposure of 100 seconds will deliver a $\lambda/\delta\lambda$=100,000 spectrum of the planet β Pictoris b with an SNR of ~25 per 1.5 km sec$^{-1}$ resolution element.

We subsequently take this value to simulate what we can learn about the atmosphere of β Pictoris b with the E-ELT using Doppler imaging techniques. Doppler imaging[26-28] can map inhomogeneities on a rotator's surface by monitoring the distortions in rotationally broadened absorption lines. By monitoring features in the velocity broadening function to appear, shift in velocity and disappear, the planet brightness distribution can be solved for using maximum entropy mapping[29], as done for stars. The reader should note that very recently, a first Doppler image map was produced for the nearby brown dwarf Luhman 16B (K=9.73) using CRIRES on the VLT, showing large-scale bright and dark features, indicative of patchy clouds[23]. The planet β Pictoris b is a factor 13 fainter than Luhman 16B, meaning that such result can be obtained almost two times faster with the E-ELT for the planet, than with the VLT for the brown dwarf. To simulate this we have constructed a toy model map of the planet's atmospheric surface of 360x180 pixels with an uniform surface brightness, except for one spot twice as bright, at a latitude of $+55^\circ$ and with a radius of $10^\circ$. Note that we gave the planet an obliquity of $30^\circ$. A low spin-axial tilt will result in a degeneracy for features on opposite hemispheres, although this is not relevant for these forward simulations. We subsequently calculated the overall spectrum of the planet for three viewing angles, which are consecutively observed at one hour intervals. The CO model that fits our CRIRES data best is assigned to each pixel in the planet surface map. Subsequently, each surface-pixel visible from the modelled viewing angle attributed to the overall spectrum, Doppler shifted according to the radial component of its rotation velocity and weighed by the cosine of the local surface viewing-angle.

The three observations were simulated with an exposure time of 15 minutes each, meaning they have a SNR of 75 per resolution element. Gaussian noise was added to the three spectra accordingly. The three spectra were subsequently cross-correlated with the model spectrum, of which the result is shown in the right panel of Extended Data Figure 2. It shows the great potential of obtaining a 2D map of the planet by using maximum entropy mapping. This can lead to unique exploration tools, such as the monitoring of multiple spots or other atmospheric features like dark or bright bands at different planet latitudes, which can solve for effects from the planet obliquity, oblateness, limb darkening, and/or differential rotation.

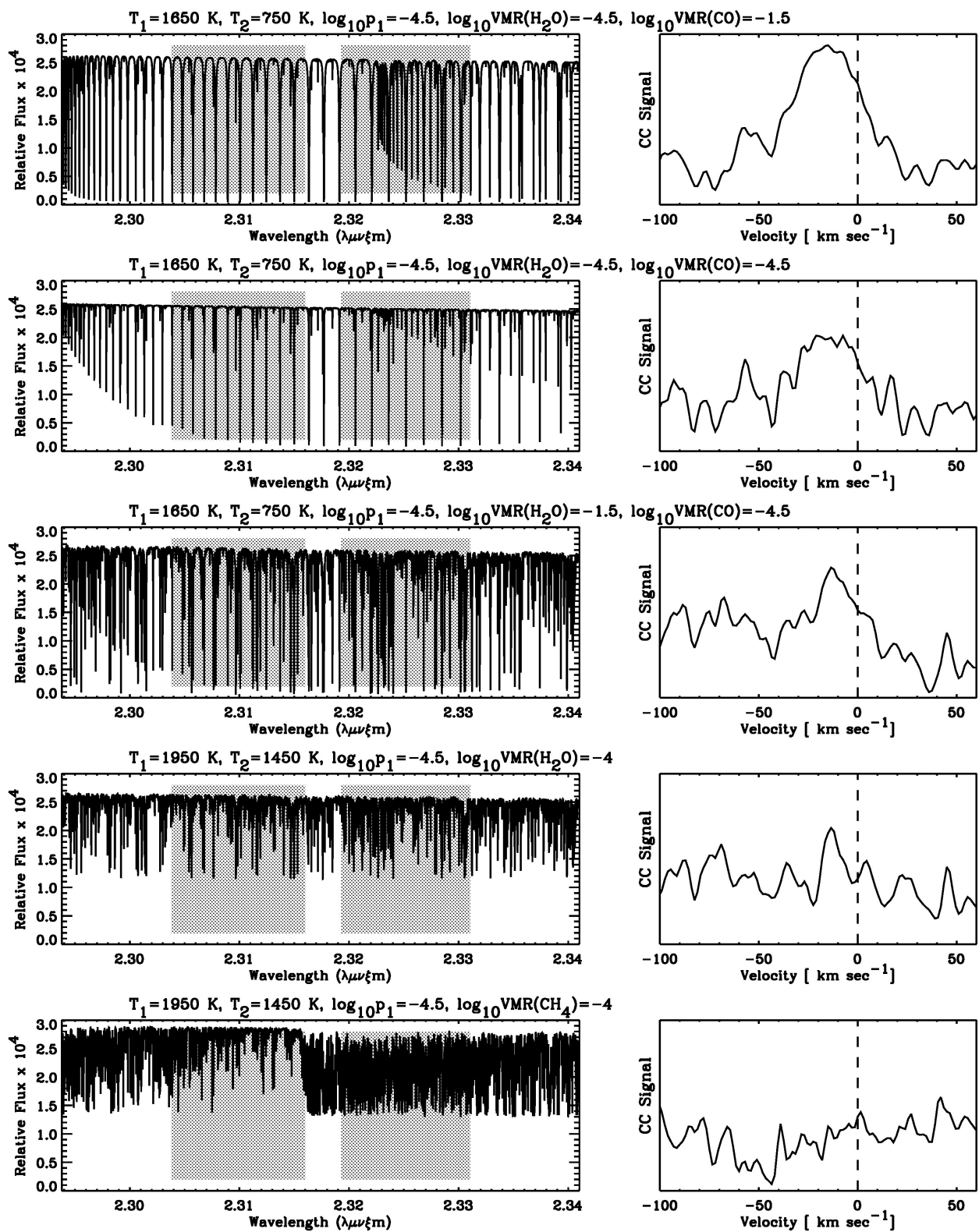

**EXTENDED DATA FIGURE 1**

**Model Spectra and cross-correlation signals.** The model spectral templates (left panels) and the resulting cross-correlation signals (right panels) for (from top to bottom), 1) a high VMR CO model, 2) a low VMR CO model, 3) the same but adding a high VMR of $H_2O$, 4) a $H_2O$ only mode, and 5) a $CH_4$ model. Atmospheric pressures are in units of bar.

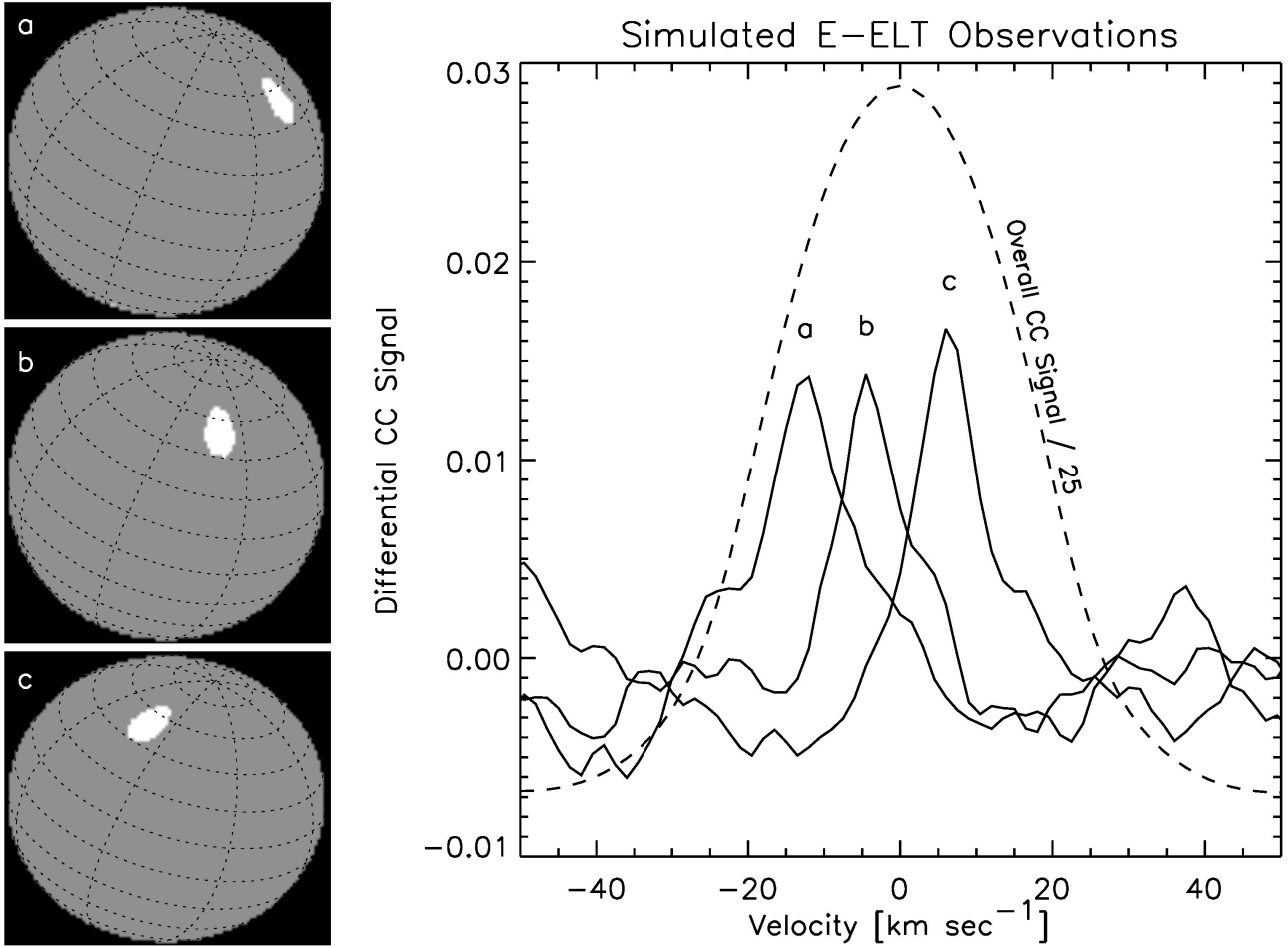

EXTENDED DATA FIGURE 2

**Simulated E-ELT observations.** We simulated observations using the future 39 m European Extremely Large Telescope (E-ELT) of a rotating spot on β Pictoris b. Such observations could be conducted with the planned METIS[40]. The three panels on the left, denoted a, b, and c, show the position of the spot approximately 1 hour apart, which was given a surface brightness of twice of the rest of the planet atmosphere. The right panel shows the difference between three cross-correlation signals with respect to the average cross-correlation profile as indicated by the dashed curve (scaled down by a factor 25), with the spot signature moving from -15 to +5 km sec$^{-1}$.